\documentclass[aps,amsmath,twocolumn,amssymb,floatfix,showpacs,superscriptaddress,nofootinbib,longbibliography]{revtex4-1}
\usepackage{mathtools}
\usepackage{braket}
\usepackage[dvipsnames]{xcolor}
\usepackage{float}
\usepackage{subfigure}
\usepackage{graphics}
\usepackage{bm}
\usepackage{soul}
\usepackage{MnSymbol}
\usepackage{braket}
\usepackage{float}
\usepackage{tikz}
\usepackage{blkarray}
\usepackage{pgffor}
\usepackage{float}
\usepackage{silence}
\usepackage[colorlinks=true,linktoc=page,linkcolor=BrickRed,citecolor=magenta,urlcolor=purple]{hyperref}

\mathchardef\mhyphen="2D 

\newcommand{\etal}{{\it et al.}}

\newcommand\bea{\begin{eqnarray}}
\newcommand\eea{\end{eqnarray}}
\newcommand\beq{\begin{equation}}  
\newcommand\eeq{\end{equation}}

\newcommand{\non}{\nonumber}  
\usepackage[normalem]{ulem}
\definecolor{lime}{HTML}{A6CE39}
\usepackage{sidecap,tikz}
\DeclareRobustCommand{\orcidicon}{\hspace{-1.0mm}
	\begin{tikzpicture}
		\draw[lime, fill=lime] (0.0,0.0) 
		circle [radius=0.15] 
		node[white] {{\fontfamily{qag}\selectfont \tiny \,ID}};
		\draw[white, fill=white] (-0.0525,0.095) 
		circle [radius=0.007];
	\end{tikzpicture}
	\hspace{-3.0mm}
}
\foreach \x in {A, ..., Z}{\expandafter\xdef\csname orcid\x\endcsname{\noexpand\href{https://orcid.org/\csname orcidauthor\x\endcsname}
		{\noexpand\orcidicon}}
}

\AtBeginDocument{%
	\newwrite\bibnotes
	\def\bibnotesext{Notes.bib}
	\immediate\openout\bibnotes=\jobname\bibnotesext
	\immediate\write\bibnotes{@CONTROL{REVTEX41Control}}
	\immediate\write\bibnotes{@CONTROL{%
			apsrev41Control,author="08",editor="1",pages="1",title="1",year="1"}}
	\if@filesw
	\immediate\write\@auxout{\string\citation{apsrev41Control}}%
	\fi
}%
\begin{document}

\title{Gate-tunable Josephson diode effect in Rashba spin-orbit coupled quantum dot junctions}

\author{Debika Debnath\orcidA{}}
\affiliation{Theoretical Physics Division, Physical Research Laboratory, Navrangpura, Ahmedabad-380009, India}
\author{Paramita Dutta\orcidB{}}
\affiliation{Theoretical Physics Division, Physical Research Laboratory, Navrangpura, Ahmedabad-380009, India}
\begin{abstract}
We theoretically explore Josephson diode effect (JDE) in superconductor/quantum dot (QD)/superconductor junction in the presence of a magnetic field and Rashba spin-orbit interaction (RSOI). We calculate the Josephson current in our QD junction using Keldysh non-equilibrium Green's function technique. We show that JDE is induced in our chiral QD junction with large rectification coefficient (RC) in the presence of RSOI and external magnetic field simultaneously. Interestingly, the sign and magnitude of the RC are highly controllable by the magnetic field and RSOI. For realistic RSOI strength in the presence of magnetic field and chirality, the RC can be tuned to be as high as $70\%$ by an external gate potential, indicating a giant JDE in our QD junction. Our proposed QD based Josephson diode (JD) may serve as a potential superconducting device component.
\end{abstract}

\maketitle
\section{Introduction}
Unidirectional operation mechanisms have made diodes an inevitable part of modern electronic devices, like current rectifiers, spintronics, quantum computers, switching devices, etc. Experimental advancements have led the path of diode formation from semiconductor diode~\cite{Baun1875} to the very recent fabrication of superconducting diode (SD)~\cite{Ando2020}. In the breakthrough experiment, Ando \etal~observed a superconducting diode effect (SDE) in a non-centrosymmetric Rashba superlattice by sandwiching Nb/V/Ta layers~\cite{Ando2020}. In the follow-up experiment, Baumgartner \etal~showed the  supercurrent rectification through the magnetochiral anisotropy in a two-dimensional electron gas (2DEG)~\cite{BaumgartnerNature}. The reason behind this evolution toward the SDE is inscribed into the directional preference of dissipationless supercurrent, while the conventional diodes are entirely based on dissipative current~\cite{Zhang2003}. Afterward, the tunability of the unidirectional supercurrent via the phase difference between superconductor leads has drawn the attention of the community and brought JDs to the forefront of research during the last couple of years~\cite{Nagaosa2021, Zhang2022, Strambini2022, Lin2022, Narita2022, Daido2022, Picoli2023, Hu2023, Souto2022, Wei2022, Davydova2022, Zinkl2022, Fominov2022, Lu2023,Banerjee2024}. The possibility of wide applications has further sharpened the questions related to achieving higher rectification with possible external tunabilities.

To realize JDE, both inversion symmetry (IS) and time-reversal symmetry (TRS) breaking have been utilized in the literature, leading to the asymmetric current-phase relation (CPR): $I_c(\Phi)\neq-I_c(-\Phi)$~\cite{Davydova2022, Martin2009,Chatterjee2023}. IS-breaking differentiates the current carried by electrons from the hole counterpart. On the other hand, breaking of the temporal symmetry ensures that the flow of up-spin electrons differs from the down-spin electrons. To break the IS, the presence of chirality~\cite{Cheng2023} or spin-orbit interaction (SOI)~\cite{Ando2020} is worth considering. For TRS breaking, one of the most convenient ways is applying an external Zeeman field, which offers additional freedom for controlling the nonreciprocal current. Otherwise, introducing intrinsic magnetism is also helpful for obtaining the JDE~\cite{Rikken2001,Nagaosa}. Thus, depending on the ways of TRS breaking, JDs are classified into two categories: (1) extrinsic JDs where the TRS is broken by external magnetic field or flux~\cite{Pal2022, LiangFu2022} and (2) intrinsic JDs with intrinsically broken TRS without any external source of TRS breaking as supported by recent works~\cite{Lin2022,Scammell2022}. In some works, JDE was explained in terms of finite momentum cooper-pairs where both TRS and IS are broken~\cite{LiangFu2022,Daido2022, Nagaosa2022b}. These findings have opened the possibility of achieving efficient JDs in various ways.

Several works have been done to show JDE considering Rashba superconductor~\cite{Bauriedl2022, Ando2020}, van der Waals heterostructure~\cite{Wu2022}, topological insulator~\cite{DanielLoss2022, Lu2023}, Dirac-semimetal~\cite{Chen2023,Pal2022}, 2DEG~\cite{Costa2023}, single magnetic atom~\cite{Martina2023}, carbon nanotube~\cite{Nagaosa2023}, InSb nanoflag~\cite{Turini2022}, normal metal~\cite{Liu2024} band asymmetric metal~\cite{abhiram2023}, topological superconductor~\cite{Cayao2023,Liu2023} etc. Not only higher dimensional systems, zero-dimensional QD has also been used to show JDE~\cite{Cheng2023,QSun_magnetic,Ortega-Taberner2023}. The context of QDs is very relevant since QDs are thoroughly explored in Josephson junctions (JJs)~\cite{QSun2000, QSun2005, Martin2007, Martin2009, Tang2016, abhiram2021,Ortega-Taberner2023} due to its potential applications to quantum information~\cite{DanielLoss1998}, spin qubits~\cite{Squid2023}, single-electron transport~\cite{Lee2014}, spintronic devices~\cite{Hirohata2022}, and also medical science and nanotechnology~\cite{Linke2023}. 

Very recently, Cheng \etal~have shown JDE in chiral QD based JJ in the presence of an external magnetic field~\cite{Cheng2023}, while Sun \etal~have studied JDE in QD based JJ with magnetic impurity~\cite{QSun_magnetic}. In the chiral QD, the chiral asymmetry can be introduced through distortion thus making it suitable for nonreciprocity~\cite{Chen2023}. However, the study of JDE in QD based junctions is very limited. An extensive study is necessary to exploit QDs meticulously for JDE or  other SDEs. Noteworthy, in Ref.~[\onlinecite{Cheng2023}], Cheng \etal~have not considered any RSOI~\cite{Rashba1984}. In reality, the presence of RSOI is anticipated in such junctions~\cite{Yokoyama2014, Rasmussen2016, Martin2007}. It not only breaks the spin-degeneracy~\cite{Jensen1996, Tokura2011}, but also the splitting due to RSOI is tunable~\cite{Takatomo1997, Dagan2010} via an external electric field, which makes it perfect for studying the current. Very recently, Mao \etal~have explored RSOI in superconducting nanowire to generate a spin diode effect~\cite{Mao2023}. 
\begin{figure}
\includegraphics[width=0.73\linewidth]{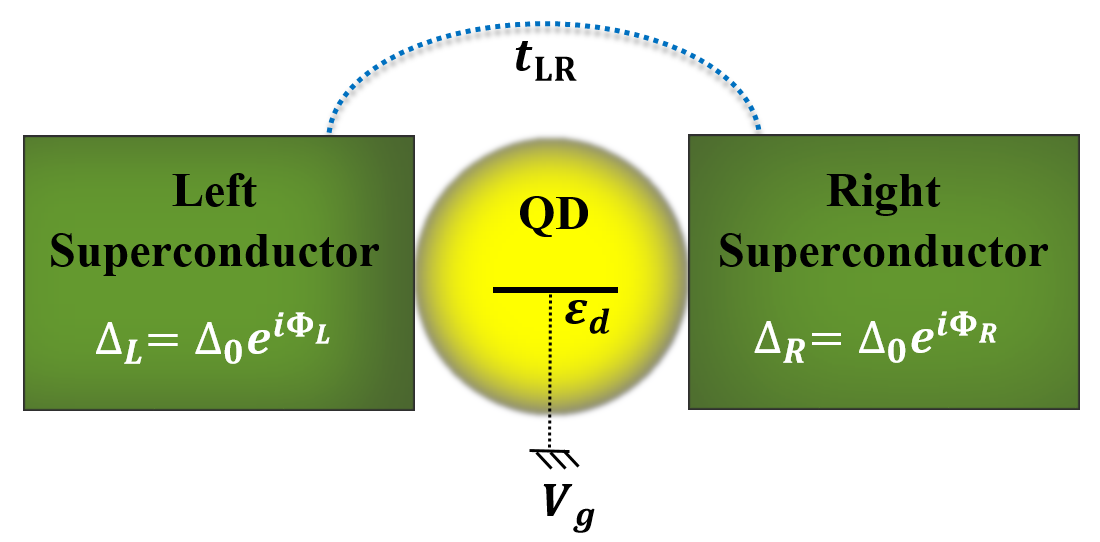}\includegraphics[width=0.18\linewidth]{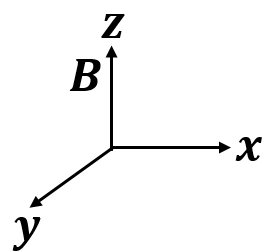}
\caption{The schematic diagram of QD-based JJ in presence of an external magnetic field and gate-voltage. A direct coupling between the superconducting leads is marked by the blue dotted line.}
\label{fig:model}
\end{figure}

With this motivation, we show JDE in a QD based JJ where a QD with RSOI is coupled to two superconductor leads in the presence of an external magnetic field. To calculate the current flowing through the junction, we employ the non-equilibrium Keldysh Green's function method~\cite{Keldysh1964} where the self-energy and the Keldysh retarded and advanced Green's functions are found using Dyson's equation of motion~\cite{Mahan}. We obtain the critical current for a range of superconducting phases, magnetic field, and RSOI. The energy of the single-level QD is controllable via an external gate-voltage~\cite{Dam2006}. We also study the effect of the gate voltages on the RC of our JD since gate-voltage can control the rectification phenomenon as shown in a recent experiment~\cite{Mohit2023}. Interestingly, the non-reciprocity of the critical currents, i.e., $I_c^+$ (positive critical current) $\ne I_c^-$ (negative critical current) in our chiral QD-based JD is enhanced compared to the previously studied non-RSOI chiral QD junction~\cite{Cheng2023}. Most interestingly, the sign and magnitude of the RC are tunable by the external magnetic field, RSOI, and the gate voltage. 

We organize the rest of the article as follows. In Sec.~\ref{Sec:model}, we present our QD-based JD model and the Keldysh Green's function formalism. We present our results and discussions in Sec.~\ref{Sec:results}. Finally, in Sec.~\ref{Sec:conclu}, we summarize our findings and conclude with some remarks.

\section{Model and Formalism}\label{Sec:model}
Our model for QD JJ is schematically shown in Fig.\,\ref{fig:model} where a QD is sandwiched between the left and right superconductor leads, denoted by L and R, respectively. An external magnetic field is applied along the direction perpendicular to the flow of the current (within the plane) as depicted in Fig.\,\ref{fig:model}. For the sake of simplification, we consider the QD consisting of a single energy level that is tunable by an external gate-voltage $V_g$. A direct coupling between the two superconducting leads is also introduced to incorporate the effect of RSOI in the single level QD junction  by virtually creating an extra path for electron transport. In the literature, this direct coupling between the leads has been extensively utilized theoretically and experimentally for decades to study the interference and tunnelling~\cite{Zacharia2001, Zeng2002, Ma2004}.
\begin{figure}
    \includegraphics[scale=0.4]{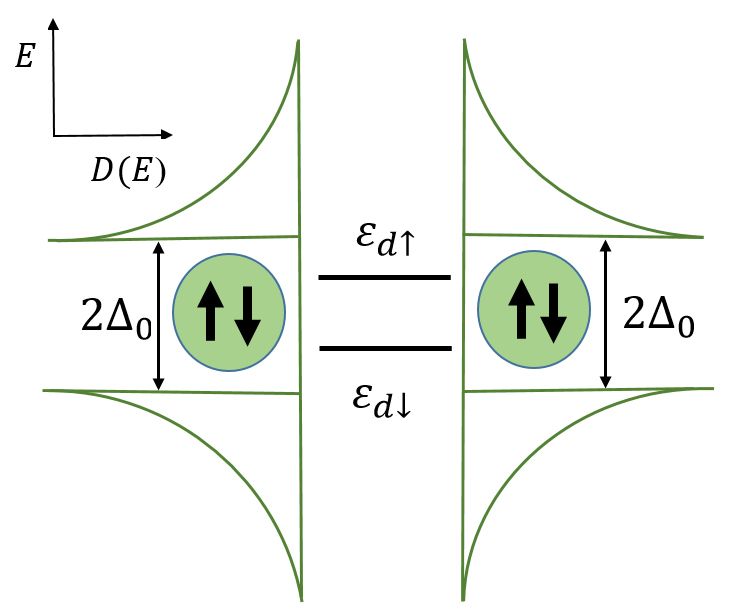}
    \caption{Density of states $D(E)$ of the superconductor leads and the energy levels of the QD.}
    \label{fig:banddiagram}
\end{figure}

We describe our QD-based JJ using the tight-binding Hamiltonian as
\begin{equation}
H=H_{\rm QD}+H_{\rm RSOI}+H_{\rm L}+H_{\rm R}+H_{\rm T}+H_{\rm LR}
\label{Ham:model}
\end{equation}
where $H_{\rm QD}$, $H_{\rm RSOI}$, $H_{\rm L(R)}$, $H_{\rm T}$ and $H_{\rm LR}$ represent the Hamiltonian for QD, RSOI, the superconducting left (right) lead, the tunneling between the QD and leads, and the direct coupling between the two leads. These individual Hamiltonians read as
\begin{eqnarray}
H_{\rm QD}&=&\sum_{\sigma}[(\varepsilon_{\rm d}-eV_{\rm g}+\sigma B+\sigma\lambda I)c^{\dagger}_{d\sigma}c_{d\sigma},
\label{ham:QD}\\
H_{\rm RSOI}&=&\alpha_{\rm R}\sum_{ d^{\prime} d}[t^{x}_{d^{'}d}(c^{\dagger}_{d^{\prime},\sigma} c_{d,\sigma}
-c^{\dagger}_{d^{'},\bar{\sigma}}c_{d,\bar{\sigma}})\nonumber\\
&&~~~~~~~~~~~~+t^{z}_{d^{'}d}(c^{\dagger}_{d^{'},\bar{\sigma}} c_{d,\sigma}-c^{\dagger}_{d,\bar{\sigma}}c_{d^{'},\sigma})],
\label{ham:RSOI}\\
H_{\rm L(R)}&=&\sum_{\alpha \in \{k{\rm L},k^{\prime}{\rm R}\}, \sigma}\varepsilon_{{\alpha}}a^{\dagger}_{{\alpha}\sigma}a_{{\alpha}\sigma}
+\sum_{\alpha\in \{k{\rm L},k^{\prime}{\rm R}\}}[\Delta_{\alpha} a_{\alpha\downarrow} a_{-\alpha\uparrow} \nonumber\\
&& ~~~~~~~~~~~~~~~~~~~~~~~~~~~~~~~~+\Delta^{*}_{\alpha}a^{\dagger}_{-{\alpha}\uparrow} a^{\dagger}_{{\alpha}\downarrow}],
\label{ham:lead}\\
H_{\rm T} &= &\sum_{\sigma} \left[\sum_{k{\rm L}}v_{\rm L}a^{\dagger}_{k{\rm L}\sigma}c_{d\sigma}+\sum_{k^{\prime}{\rm R}}v_{\rm R}a^{\dagger}_{k^{\prime}{\rm R}\sigma}c_{d\sigma}+\text{h.c.}\right], \nonumber \\
\label{ham:T}\\
H_{\rm LR}&=&t_{\rm LR} (a^{\dagger}_{k{\rm L}\sigma}a_{k^{\prime}{\rm R}\sigma}+a^{\dagger}_{k^{\prime}{\rm R}\sigma}a_{k{\rm L}\sigma}).
\label{ham:LR}
\end{eqnarray}
In Eq.\,\eqref{ham:QD}, $\varepsilon_{\rm d}$ represents the energy level of the QD, which is tunable by the external voltage $V_{\rm g}$ (shown in Fig.~\ref{fig:model}), B is the external magnetic field applied along $z$-direction. Following Lenz's law, introducing the chirality in the QD generates an induced magnetic field which is proportional to the current flowing through the dot. The fourth term of Eq.\,\eqref{ham:QD} is responsible for the induced magnetic field with $\lambda$ being the proportionality constant~\cite{Cheng2023}. The Pauli matrix $\sigma$ acts on the spin degree of freedom. $c^{\dagger}_{d\sigma}(c_{d\sigma})$ is the creation (annihilation) operator for the electrons in the QD where the notation $d$ is used to identify the dot.

In Eq.\,\eqref{ham:RSOI}, $\alpha_{\rm R}$ is the strength of the RSOI. Due to the RSOI, the single energy level of the QD is split in two levels as shown in Fig.~\ref{fig:banddiagram}. Considering two states $d$ and $d^{'}$ in the QD, the first term of Eq.\,\eqref{ham:RSOI} represents the hopping of same spin electrons between two different states and the second term corresponds to the spin-flip hopping. The second quantized form of the RSOI is derived~\cite{Kuntal2023} from the Hamiltonian given by,
\bea
H_{\rm RSOI}&=&\hat{y}\cdot \frac{\alpha_{\rm R}}{\hbar}\left[\mathbf{\sigma} \times (\mathbf{p}+\frac{e\mathbf{A}}{c})\right]
\eea
where $\bf{A}$ is the magnetic vector potential and $\bf{p}$ is the momentum of electrons. Using the gauge $(0, Bx,0)$, we can express the hopping as
$t^{x(z)}_{d^{'}d}=\int dr \psi^{*}_{d^{'}}(r)p_{x(z)}\psi_{d}(r)$
where $\psi_d(r)$ is the orbital wavefunction for the QD electrons.

In Eq.\,\eqref{ham:lead}, $\varepsilon_{\alpha}$ with $\alpha\in \{k{\rm L}, k^{\prime}{\rm R}\}$ denotes the onsite energy of electrons in both leads, $a^{\dagger}_{k {\rm L}(k^{\prime}{\rm R})\sigma}(a_{k{\rm L}(k^{\prime}{\rm R})\sigma})$ is the creation (annihilation) operator of the electrons in the left (right) lead with momentum $k{\rm L}$ ($k^{\prime}{\rm R}$). The superconducting pair potential is denoted by: $\Delta_{\rm L(R)}=\Delta_{\rm 0} e^{i\Phi_{\rm L(R)}}$ where $\Phi_{\rm L}$ ($\Phi_{\rm R}$) is the superconducting phase for the left (right) lead. 
The coupling strength between the left (right) superconductor and the QD is described by $v_{\rm L}(v_{\rm R})$ in Eq.\,\eqref{ham:T} and the direct tunneling coefficient between the two leads is represented by $t_{\rm LR}$ in Eq.\,\eqref{ham:LR}~\cite{QSun2005, Kuntal2023}. 

In order to simplify the total Hamiltonian, we now apply two unitary transformations. In one transformation, the superconducting phase $\Phi_{\rm L(R)}$ and energy gap $\Delta_{\rm 0}$ will be decoupled using the generator $U_1$, while in the other transformation based on the generator $U_2$, the effect of RSOI will be included in the tunneling current. Hence, the two consecutive unitary transformations are applied to Eq.\,\eqref{Ham:model} with the following generators: 
\begin{eqnarray}
     U_{\rm 1}= exp [\sum_{\rm kL(k^{\prime}R)\sigma} \frac{i\Phi_{\rm L(R)}}{2} a^{\dagger}_{\rm kL(k^{\prime}R),\sigma}a_{\rm kL(k^{\prime}R),\sigma}],\\
     U_{\rm 2}=\begin{cases}
    1 & (x<x_{\rm L})\\
    \frac{1}{\sqrt{2}} e^{-i k_{\rm R}(x-x_{\rm L})\sigma} & ( x_{\rm L}<x< x_{\rm R})\\
    \frac{1}{\sqrt{2}} e^{-i k_{\rm R}(x_{\rm R}-x_{\rm L})\sigma} & ( x_{\rm R}< x)
\end{cases}   
\end{eqnarray}
where $ k_{\rm R}=\alpha_{\rm R}\frac{m^{*}}{\hbar^{2}}$. The transformed Hamiltonian reads as
\begin{equation}
  \tilde{H}=e^{U_{\rm 2}}e^{U_{\rm 1}}He^{-U_{\rm 1}}e^{-U_{\rm 2}}.  
\end{equation}
The interlevel hopping and the spin-flip terms are neglected since our QD has a single energy level~\cite{QSun2005, Kuntal2023}. We obtain the transformed Hamiltonian as,
\begin{widetext}
\bea
\tilde{H}&=&\sum_{\alpha \in \{k{\rm L},k^{\prime}{\rm R}\},\sigma}\left[\varepsilon_{{\alpha}}a^{\dagger}_{{\alpha}\sigma}a_{{\alpha}\sigma}+\Delta_{0}a_{{\alpha}\downarrow}a_{-{\alpha}\uparrow} +
\Delta^{*}_{0}a^{\dagger}_{-{\alpha}\uparrow}a^{\dagger}_{{\alpha}\downarrow}\right]+
\left[\sum_{k{\rm L},\sigma} v_{\rm L}e^{\frac{i\Phi_{\rm L}}{2}}a^{\dagger}_{k{\rm L}\sigma}c_{d\sigma}+
\sum_{k^{\prime}{\rm R},\sigma}v_{\rm R}e^{\frac{i\Phi_{\rm R}}{2}}a^{\dagger}_{k^{\prime}{\rm R}\sigma}c_{d\sigma}e^{-i\sigma\phi_{\rm RS}}+\text{h.c.}\right] \nonumber \\
&& ~~~~~~~~~~~~~~~~~~~~~~~~~~~~~~~~~~~~~~~~~~~~~~~~~~~~~~~~~~~~~~~~~~~~~~~~~~~~~~~~~~~~~~~~~~~~~~~~~~~~~~~~~~+H_{\rm QD}+H_{\rm LR}.
\label{Eq:H_eff}
\eea
\end{widetext}
Notably, the overall RSOI strength is now represented by the RSOI-induced phase factor $\phi_{\rm RS}$ which appears in the tunneling part of the transformed Hamiltonian and it is related to the $\alpha_{\rm R}$ by the expression, $\phi_{\rm RS}=\alpha_{\rm R}\frac{m^{*}}{\hbar^{2}} l$, where $ l$ is the length scale of RSOI. 

With this transformed Hamiltonian, the Josephson current in our QD-based junction can be found using the relation~\cite{Meir1994, Datta1997} 
\begin{equation}
I=-e \left<\frac{dN_{\rm L(R)}}{dt}\right> =ie \left<\left[N_{\rm L(R)},\tilde{H}\right]\right>
\end{equation}
where the number operator is given by $N_{\rm L}=\sum\limits_{\rm k{\rm L},\sigma} a^{\dagger}_{k{\rm L}\sigma} a_{k {\rm L}\sigma}$. To calculate the current, we use the Keldysh non-equilibrium Green's function formalism which is one of the most efficient techniques to solve quantum transport problems. For the Keldysh lesser Green's functions, we use the fluctuation-dissipation theorem~\cite{Kubo, kamenev_2011} and obtain the Josephson current expression as,
\begin{eqnarray}
I&=&\frac{e}{\pi}\int d\epsilon \sum_{\sigma} {\rm Re}[v_{\rm L} e^{\frac{i\Phi_{\rm L}}{2}} \{G^{<}_{\rm dL,11}(\epsilon)+G^{<}_{\rm dL,33}(\epsilon)\}\nonumber\\
&&~~~~~~~~~~~~+t_{\rm LR} \{G^{<}_{\rm RL,11}(\epsilon)+G^{<}_{\rm RL,33}(\epsilon)\}]
\label{eq:JC}
\end{eqnarray}
where $G^{<}_{ii}(\epsilon)$ is the $ii-$th element of the Fourier transformed time-dependent Keldysh Green's function $G^{<}(t)$. 
\begin{figure*}
    \includegraphics[scale=0.68]{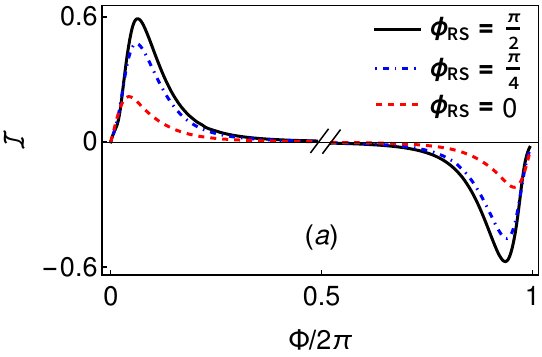}
    \vskip 0.4cm
     \includegraphics[scale=0.4]{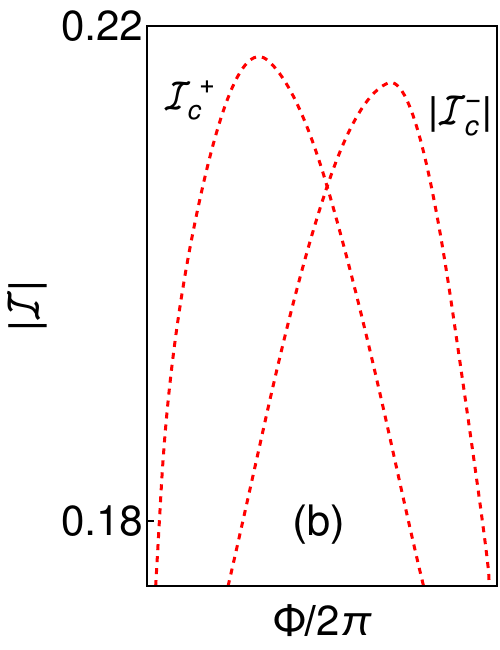}~~~~~~~
      \includegraphics[scale=0.4]{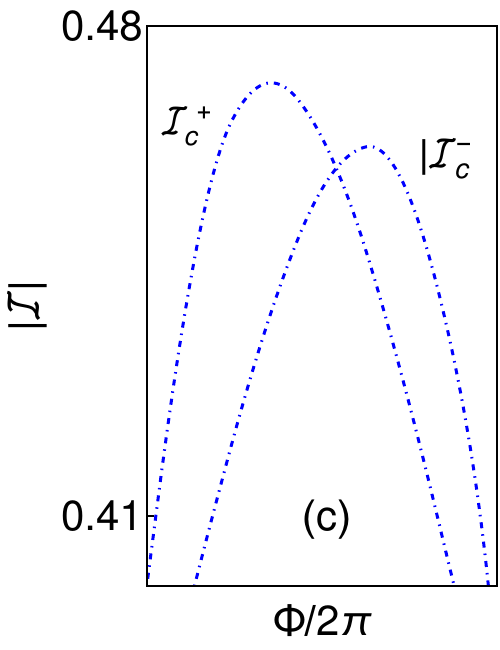}~~~~~~
       \includegraphics[scale=0.4]{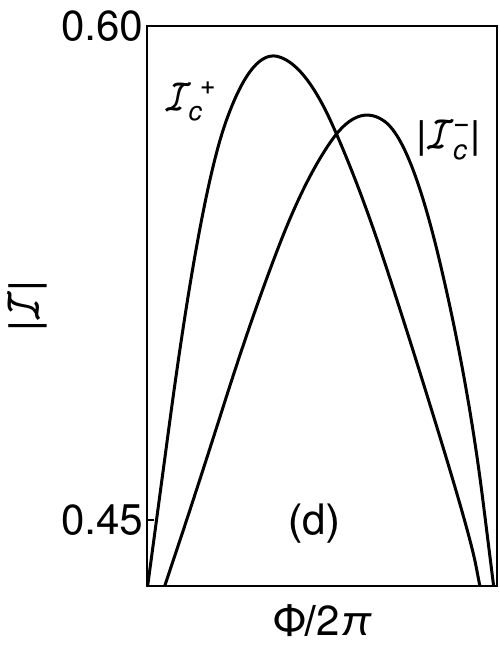}
    \caption{(a) Josephson current ($I$) in units of $e\Delta_{0}$ as a function of superconducting phase difference $(\Phi$) at $B=0.6$ with and without Rashba phase. (b) - (d): A comparison between $I_c^+$ and $|I_c^-|$ for $\Phi/2\pi$ values along the $x$-axis given by (b) $I_c^+$: $0.02 \le \Phi/2\pi \le 0.07$ and $|I_c^-|$: $0.75\le \Phi/2\pi \le 0.78$, (c) $I_c^+$: $0.03 \le \Phi/2\pi \le 0.07$ and $|I_c^-|$: $0.72\le \Phi/2\pi \le 0.76$, and (d) $I_c^+$: $0.03 \le \Phi/2\pi \le 0.08$ and $|I_c^-|$: $0.71\le \Phi/2\pi \le 0.75$,  respectively.}
    \label{fig:Jvsphi_R}
\end{figure*}

For our model Hamiltonian in Eq.\,\eqref{Ham:model}, the Keldysh lesser Green's function can be written as,
\begin{eqnarray}
G^{<}=\left(\begin{array}{ccc}
G_{\rm LL}^{<}&G_{\rm LR}^{<}&G_{\rm Ld}^{<}\\
G_{\rm RL}^{<}&G_{\rm RR}^{<}&G_{\rm Rd}^{<}\\
G_{\rm dL}^{<}&G_{d \rm R}^{<}&G_{\rm dd}^{<}
\end{array}\right)
\label{Greenless}
\end{eqnarray}
which is a $12 \times 12$ matrix in the spin $\bigotimes$ Nambu basis and it follows $G^{<}_{\rm RL}=-(G^{<}_{\rm LR})^*$. 
Moreover, the Keldysh retarded Green's function can be expressed in terms of the self-energy using the Dyson equation of motion~\cite{Mahan}:
$G^{\rm r}=g^{\rm r}+g^{\rm r}\Sigma^{\rm r}G^{\rm r}$,
where $g^{\rm r}$ is the Keldysh retarded Green's function for the uncoupled QD and leads. The self-energy $\Sigma^r$ carries all the information about the tunnelings. We refer to Appendix~\ref{apnd1:cal} for the details of the self-energy calculations. Finally, using Eq.\,\eqref{eq:JC}, we calculate the Josephson current self-consistently considering the effect of the induced field. 

To quantify the quality of the rectification by our JD, we define the diode RC as,
\begin{eqnarray}
\mathcal{R}=\frac{I_c^+-|I_c^-|}{I_m} \times 100\%  
\label{RC}
\end{eqnarray}
where we normalize the rectificaion by the mean current given by $I_m=(I_c^++|I_c^-|)/2$.

Throughout the rest of the manuscript, we consider the superconducting phases in the two leads as $\Phi_{\rm L}=\Phi/2$ and $\Phi_{\rm R}=-\Phi/2$ so that the phase difference becomes $\Phi_{\rm L}-\Phi_{\rm R}=\Phi$. We use the natural units where $m^{*}=1$ and $\hbar=1$, and calculate the Josephson current $I$ in units of $e\Delta_{0}$. Also, we set $v_{\rm L}=v_{\rm R}=0.5\Delta_0$, $\lambda=0.05$, $t_{\rm LR}=0.2\Delta_0$, and $eV_{\rm g}=0$ unless specified. We discuss the effects due to the change in the parameter values in the upcoming section. It is important to mention that we consider the magnetic field only along the perpendicular direction to maximize the rectification following the discussions in some recent works~\cite{Bauriedl2022, Lotfizadeh2023,Cheng2023}. We show all the results only for symmetric leads where the pair potential is the same for both leads. A detailed discussion on the effects of asymmetric leads can be found in Appendix~\ref{apnd2:asymmetric}. 


\section{Results and Discussions}\label{Sec:results}
In this section, we present and discuss the behaviors of the current followed by the RC for the various parameters considered in our model. 

\subsection{Current-phase relation}
We begin by referring to Fig.~{\ref{fig:Jvsphi_R}}(a), where we show the CPR at a particular magnetic field for various Rashba phases. The phase difference between the two leads establishes a Josephson current with the CPR $I=I_c \sin{\Phi}$ similar to an ordinary JJ as the Cooper pair tunnels through the channels of the QD (see Fig.~\ref{fig:banddiagram}). The Josephson current in our junction also follows $I(n\pi)=0$ where $n=\pm1,\pm2,...$ and so on. For a better understanding of the current profiles, we zoom around the peaks and skip showing the current around $\Phi/(2\pi)=0.5$. In order to compare the magnitudes, we again show the positive ($I_c^{+}$) and the negative ($|I_c^{-}|$) critical currents separately in Fig.~{\ref{fig:Jvsphi_R}}(b-d) for Rashba phases $\phi_{\rm RS}=0$, $\pi/4$, and $\pi/2$, respectively. In Fig.~\ref{fig:Jvsphi_R}(b), we observe that the positive critical current $I_c^+={\rm max}[I(0<\Phi<\pi)]$ is not the same as the negative critical current $I_c^-={\rm max}[-I(\pi<\Phi<2\pi)]$ i.e., $I_c^+\ne |I_c^-|$. This describes that for a particular phase difference, the forward current is different from the reverse current, indicating a JDE. At $\phi_{\rm RS}=0$, the JDE happens in the QD junction due to the presence of the external ($B\ne0$) and induced field ($\lambda\ne0$) that simultaneously break the TRS and chiral symmetry, respectively, similar to the results presented in Ref.~\cite{Cheng2023}. However, the difference between $I_c^+$ and $|I_c^-|$, where $I_c^+>|I_c^-|$, is very low in the absence of RSOI (see Fig.~{\ref{fig:Jvsphi_R}}(b)).

Now, with finite RSOI ($\phi_{\rm RS}\ne0$), the magnitudes of the critical currents and also the asymmetry between them increases as depicted in Fig.~\ref{fig:Jvsphi_R}. The inclusion of RSOI breaks the IS, which results in the enhancement of the nonreciprocity of the current, and thus, the diode effect increases. Here, we want to make a note that in the absence of any chirality ($\lambda=0$) and magnetic field ($B=0$), the CPR in a RSOI coupled QD junction follows the analytical form $I\simeq v_{\rm L} \cos{\frac{\phi_{\rm RS}}{2}} \left[\sin\left(\frac{\Phi+\phi_{\rm RS}}{2}\right)+\sin\left(\frac{\Phi-\phi_{\rm RS}}{2}\right)\right]$ (see Appendix~\ref{apnd:currexpression} for the derivation). We refer to Appendix~\ref{apnd3:zerochirality} for the results with $\lambda=0$. Also, we show all the results for one particular coupling strength ($v_{\rm L}=v_{\rm R}$). Any change in the coupling strength will not affect the qualitative behavior of the current in our junction as long as it is less than the magnetic field.

\begin{figure}
\includegraphics[width=7.2cm,height=4.8cm]{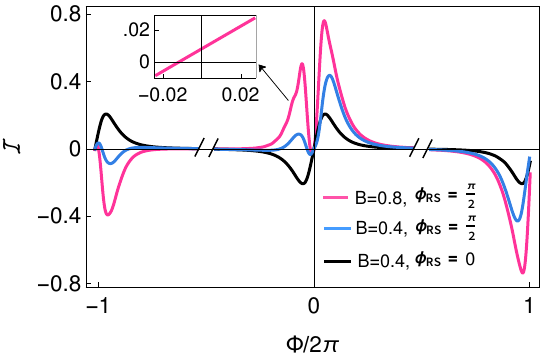}
\caption{Josephson current $(I)$ in units of $e\Delta_{0}$ as a function of superconducting phase difference ($\Phi$). Inset: Current profile around $\Phi=0$.}
\label{fig:Jvsphi_BR}
\end{figure}

In order to investigate the effect of the magnetic field and RSOI in more detail, we now plot the same for various magnetic fields in the absence and presence of RSOI in Fig.~{\ref{fig:Jvsphi_BR}}. We see that with the increase in the magnetic field, the critical currents and also the asymmetry between $I_c^+$ and $|I_c^-|$ increase. The behavior of the enhancement of the current with the increasing magnetic field is also found in the absence of RSOI~\cite{Cheng2023}.
In an ordinary JJ, the CPR follows $I(\Phi)= -I(-\Phi)$. Interestingly, in Fig.~{\ref{fig:Jvsphi_BR}}, we observe that this relation breaks down i.e, $I(\Phi)\neq -I(-\Phi)$. We also find that in contrast to the usual $2\pi$ periodicity i.e., $I_c(\Phi+2\pi)=I_c(\Phi)$ found in an ordinary JJ, the symmetry around $\Phi=0$ is lost in the presence of magnetic field and RSOI in the QD junction, resulting in $4\pi$ periodicity of the current. A small finite current is found to be present even at the zero superconducting phase difference, i.e., $I(\Phi=0)\ne 0$ (see inset of Fig.~\ref{fig:Jvsphi_BR}) where in an ordinary JJ, the CPR follows $I(\Phi=0)=0$. This indicates the appearance of an additional phase $\Phi_0$ in the CPR and is referred to as the anomalous Josephson current in the literature~\cite{Mayer2020, Davydova2022, Martin2009, Yokoyama2014, Assouline2019}. The shifting of $I(\Phi=\Phi_{0})=0$ is also known as $\Phi_{0}-$JJ. The combined effect of the IS and the TRS breaking is reponsible for this anomalous supercurrent or equivalently the phase-shift in our junction. We check that the shifting of the $\Phi_0$ phase intensifies with the external magnetic field following $B/v_{\rm L(R)}>1$~\cite{Buzdin2005}. This anomalous behavior of the current is explained in the literature in terms of the spontaneous breaking of TRS at $\Phi=0$~\cite{Buzdin2008, Krive2005, Reynoso2008, Martin2009, Yokoyama2014}.

\subsection{Rectification without any gate voltage}
\begin{figure}
    \includegraphics[scale=0.75]{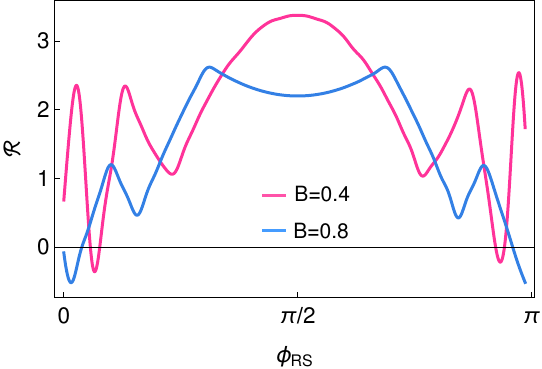}
    \caption{Rectification coefficient $(\mathcal{R}\%$ vs.\,Rashba phase $(\phi_{\rm RS})$.}
    \label{fig:QvsR_B}
\end{figure}

Till now, we see that the phenomenon of non-reciprocity follows $I_c^+>I_c^-$. It indicates that the forward current is preferred over the reverse one for all the parameter values considered so far. To analyze the rectification quality, we present the behavior of the RC using Eq.\,{\eqref{RC}} for our QD-based JD as a function of $\phi_{\rm RS}$ in Fig.~{\ref{fig:QvsR_B}}. On the whole, RC of our JD initially increases and then decreases with the increase in Rashba phase. However, the detailed behavior of the RC depends on the magnitude of $B$. There appears an interplay between the RSOI strength and magnetic field for determining the value of RC. For illustration, at $B=0.8$, the current in our QD-based JD increases as $\phi_{\rm RS}$ increases from $0$ to $\pi/2$. As $\phi_{\rm RS}$ increases further, the RC reduces following a partial mirror symmetry with respect to $\phi_{\rm RS}=\pi/2$. For a lower magnetic field strength like $B=0.4$, we observe a saddle near $\phi_{\rm RS}=\pi/2$. So, it is evident that the correlation between the $B$ and $\phi_{\rm RS}$ plays a vital role in determining the diode effect qualitatively and also quantitatively. Additionally, two more important observations are in order. The RC is low in this parameter regime. Interestingly, there exists a certain regime of RSOI, where we find negative values of the RC, which indicates that the backward current is preferred more than the forward one.

\begin{figure*}
    \includegraphics[width=0.9\linewidth]{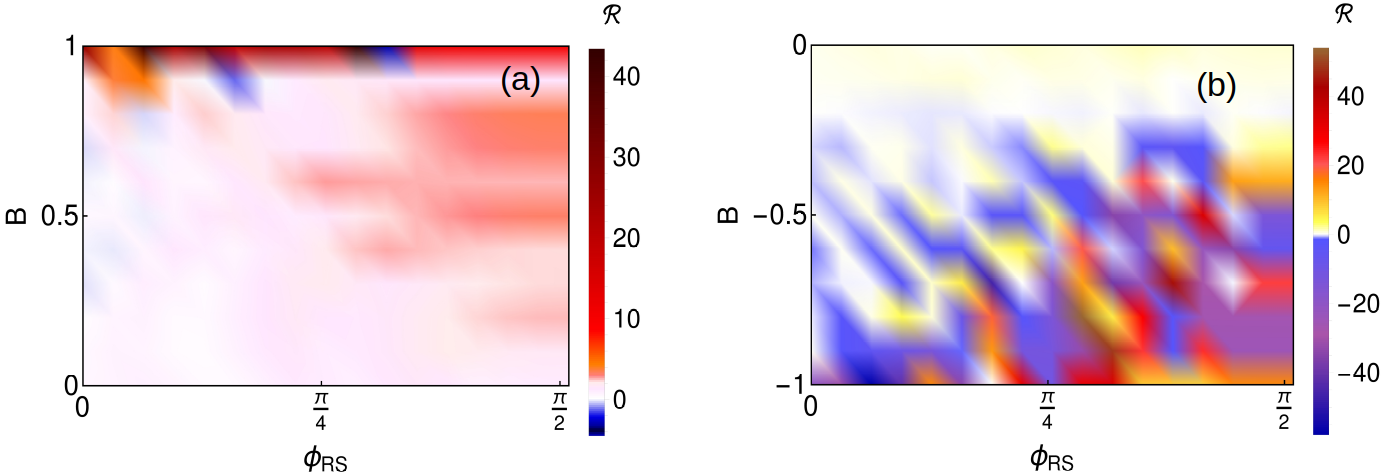}
    \caption{Density plot of the rectification coefficient $(\mathcal{R\%})$ as a function of the Rashba phase $(\phi_{\rm RS})$ and magnetic field $(B)$ applied along (a) $+z$ and (b) $-z$-direction.}
    \label{fig:Densityplot}
\end{figure*}

In order to understand the competitive effect of the RSOI and magnetic field on the diode quality, we now study the RC for broad parameter regimes of the magnetic field ($0\le B \le 1$ in panel (a) and $-1\le B \le 0$ in panel (b)) in the presence of RSOI by presenting a density plot of the RC in Fig.~\ref{fig:Densityplot}. Our observations are manifold. (i) The RC is finite but very low for $B=0$. This is expected since external TRS breaking is not necessary for the chiral QD JDE~\cite{Nagaosa2023,Lin2022,Scammell2022,Wu2022} but the effective magnetic field can not induce sufficiently large JDE in our RSOI coupled junction. (ii) RC varies with the strength of the magnetic field. (iii) The diode effect is sensitive to the direction of the applied magnetic field, i.e., $\mathcal{R}(B)\neq -\mathcal{R}(-B)$ when the directions of Rashba field and the supercurrent are unchanged. This is similar to the experimental results reported in Ref.~[\onlinecite{Bauriedl2022, Lotfizadeh2023}]. The competitive effect of the magnetic field and the RSOI on the RC is more substantial when the magnetic field is applied along $-z$-direction. (iv) Most interestingly, not only the magnitude of RC, tuning the magnetic field and RSOI simultaneously, we can also tune the RC from the positive to negative and vice-versa. Comparing the two panels of Fig.~\ref{fig:Densityplot}, we observe that when the magnetic field is applied along $+z$-direction, higher values of the RC are achieved for strong magnetic fields only, whereas the RC is amplified for a relatively much weaker magnetic field for the same RSOI when it is applied along the opposite direction. (v) Overall, we find that maximum $\mathcal{R}\sim42\%$ for $+B$ and $\sim57.5\%$ for $-B$ can be achieved in our QD-based JD. To be noted, we have checked that in the absence of any chirality ($\lambda=0$), the diode RC can be as high as $43\%$ due to only RSOI. 

We explain the observations as follows. Introduction of the RSOI splits the QD energy level. When we simultaneously switch on the external magnetic field, it extends the energy level splitting considering the spins of the electrons. Therefore, the combined effect of the magnetic field and the Rashba generates several possibilities for the tunnelling of the Cooper pairs from the left to the right superconducting lead. Hence, the correlation among both interactions results in ample possibilities or variations in the RC of our JD. It turns out that the direction of the supercurrent is strongly favored if the three vectors i.e., supercurrent, magnetic field, and RSOI lie along the Cartesian coordinates (clock-wise), which happens in the $+B$ direction in our set-up. For the $-B$, there are more fluctuations in the current with the sign of RC being sensitive to the strength of the magnetic field for any RSOI. It is essential to mention that the qualitative nature of the RC with the change in the direction of the external magnetic field from $-z$ to $+z$ matches well with the experimental study on SDE on epitaxial Al-InAs JJ, where the JDE due to strong RSOI is studied in the presence of an external magnetic field~\cite{Lotfizadeh2023}.

Before we proceed further, we comment on the direct coupling.  In our JJ model, we consider a single energy level in the QD. For the single energy level, an external direct coupling between the two leads via $t_{\rm LR}$ is necessary to include the Rashba effect in our system \cite{QSun2005, Martin2009}. This coupling opens up a pseudo-channel for the electron to tunnel. In reality, since a single QD is a very small region, it is highly likely to have a direct coupling between the two leads. Eventually, the strength of this direct hopping has a significant effect on the RC, which we discuss in Appendix~\ref{apnd4:directcoupling}.

\subsection{Gate-tunability of the diode effect}

\begin{figure}
    \includegraphics[width=0.85\linewidth]{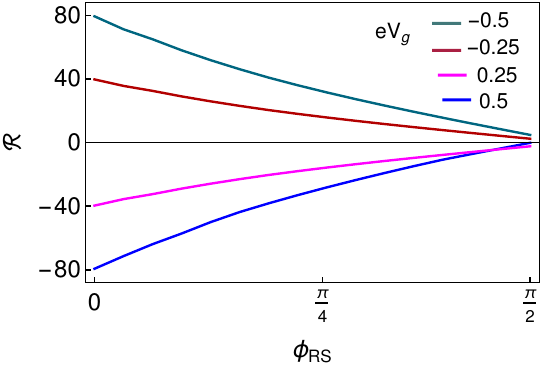}
    \caption{Rectification coefficient ($\mathcal{R}\%$) vs. Rashba phase ($\phi_{RS}$) at $B=0.4$.}
    \label{fig:QvsR_Vg}
\end{figure}

Till now, all results are shown only for the gate voltage $eV_g=0$. In Fig.~\ref{fig:QvsR_Vg}, we show the RC as a function of the Rashba phase for various gate voltages. The applied gate voltage tunes the QD energy levels. We find that by only tuning the gate voltage, it is possible to get the RC as high as $80\%$ even in the absence of any RSOI in the chiral QD. For the finite RSOI, the gate-voltage induces an effective RSOI in the system since it acts as an electric field~\cite{Hong2004,Iorio2018}. With the increase in the gate voltage, the induced RSOI through the external field becomes more effective compared to the RSOI in the ungated junctions. With the increasing Rashba phase, the RC gradually decreases for all finite gate voltages. Keeping in mind the inevitable presence of RSOI in such QD-based junction, it is possible to achieve RC $\sim70\%$ by tuning $V_{\rm g}$ for an optimum strength of $\phi_{\rm RS}$ in our proposed QD based JD, which is the highest RC of a JD reported so far for the non-transparent junction to the best of our knowledge. Also, by tuning the gate voltage from the positive to the negative values, or vice versa, we can determine the direction of the preferential current flow in our JD as shown in Fig.~\ref{fig:QvsR_Vg}. We find, at a particular RSOI, $\mathcal{R}(eV_{g})\simeq-\mathcal{R}(-eV_{g})$, i.e. the RC is almost symmetric with respect to the zero gate voltage at any RSOI.

\section{Summary and conclusion}\label{Sec:conclu}
To summarize, we have studied the JDE in single energy level chiral QD based junction. To calculate the Josephson current, we have employed the Keldysh non-equilibrium Green's function technique. We have found that the critical current in our JJ increases with the increase in RSOI strength. We show the JDE in our chiral QD based JJ in the presence of RSOI and a magnetic field. Noteworthy, in our RSOI coupled QD junction, simultaneous presence of the chirality and the magnetic field are not mandatory to achieve JDE, but they can enhance the rectification. In the presence of RSOI, the magnitude of the RC gets amplified and most importantly, it results in a sign-changing behavior of the RC. Finally, we have shown that by tuning the QD energy level via an external gate potential, it is possible to get giant JDE with $\mathcal{R}\sim70\%$ at an optimal strength of RSOI in the presence of magnetic field. 

From the experimental point of view, an estimation of the induced current is in order. Using the Biot-Savart law, the induced magnetic field is found to be $B^{'}=\frac{\mu_{0}I}{2r}$ where $r$ is the radius of the single turn of a current-carrying helix. This induced magnetic field is proportional to the current following $B^{'}=\lambda I$. Hence, comparing both the relations, we obtain $\lambda=\frac{\mu_{0}}{2r}$. Considering the radius $r=1$ nm, the induced field will be of the order of $10^{-3}$ Oe for a current $I=0.5$ nA. However, in our work, the RSOI is playing the main role for inducing the larger diode effect. The RSOI changes as per the material property. In the widely used materials for QDs like GaAs or InAs, the RSOI strength is found to be very small $\sim0.04\times 10^{-11}$ eVm and $0.28\times10^{-11}$ eVm, respectively~\cite{Eldridge2007, Miller2003, Chen1993, Hong2004}, whereas in InSb, RSOI is relatively higher $\sim 1.16 \times 10^{-11}$ eVm~\cite{Greene1992}. Therefore, for the semiconductor QD with low RSOI strength, the gate voltage could be tuned easily to achieve a very high RC, as mentioned above. Our present results support three major experimental works on the SDE of recent times: i) it reconfirms Ref.~[\onlinecite{Ando2020, BaumgartnerNature}], the breaking of IS through Rashba materials can induce the JDE, ii) by tuning the gate voltage~\cite{Mohit2023}, the RC of the JD can be increased by a large amount, iii) even in the absence of an external magnetic field it is possible to have a weak Josephson diode effect~\cite{Wu2022,Lin2022,Scammell2022}. All these newly verified results are now explored in our QD junction.

Finally, we emphasize that including RSOI in modern quantum systems can be supremely effective in studying more efficient JDE. Our RSOI coupled QD junction explains the JDE theoretically, in a more general way, with a highly efficient possibility of experimental fabrication since RSOI is unavoidable in such low dimensional junctions~\cite{Duine2015}. The tunability of the external magnetic field, the superconducting phase difference, gate voltage, RSOI and applications to a wide range of quantum materials and correlated systems, like spin qubits, spin-transistors, single electron transistors, topological materials, Dirac semi-metals, superconductors, and many more~\cite{Hirohata2022} make our proposed QD based JDs highly efficient for switching devices.


\acknowledgments{We thank Mohit Randeria, Navinder Singh, and Alireza Qaiumzadeh for helpful discussions. We acknowledge the Department of Space, Government of India for all support at PRL. P.\,D. acknowledges the Department of Science and Technology (DST), India, for the financial support through the SERB Start-up Research Grant (File no.\,SRG/2022/001121).}
 
\begin{appendix}
\appendix
\section{Calculations of Keldysh Green's function and the transport matrix}\label{apnd1:cal}

In this section, we summarize some major steps of the calculation of Keldysh Green's function and the transport matrix. 

The self-energy included in the Dyson equation of motion for the Keldysh retarded Green's function is expressed in the basis $\{\epsilon_{\uparrow},-\epsilon_{\uparrow},\epsilon_{\downarrow},-\epsilon_{\downarrow}\}$ as~\cite{QSun2005},
\begin{eqnarray}
\Sigma^{\rm r}=\left(\begin{array}{ccc}
0  &  t_{\rm LR} & V_{\rm L}\\
t^{*}_{\rm LR}  & 0 &  V_{\rm R}\\
V^{*}_{\rm L}  &  V^{*}_{\rm R}  &   0
\end{array}\right)
\label{selfenergy}
\end{eqnarray}
where
\begin{eqnarray}
t_{\rm LR}=\left(\begin{array}{cccc}
t_{\epsilon\uparrow,\epsilon\uparrow}  &  0 & 0 & 0\\
0  & t_{-\epsilon\uparrow,-\epsilon\uparrow} &  0 & 0\\
0  &  0 & t_{\epsilon\downarrow,\epsilon\downarrow}  &   0 \\
0  &  0 & 0 &  t_{-\epsilon\downarrow,-\epsilon\downarrow} 
\end{array}\right),
\label{self}
\end{eqnarray}
\begin{eqnarray}
V_{\rm L}=\left(\begin{array}{cccc}
v_{\rm L}e^{i\frac{\Phi_{\rm L}}{2}}  &  0 & 0 & 0\\
0  & -v_{\rm L}e^{-i\frac{\Phi_{\rm L}}{2}} &  0 & 0\\
0  &  0 & v_{\rm L}e^{i\frac{\Phi_{\rm L}}{2}}  &   0 \\
0  &  0 & 0 &  -v_{\rm L}e^{-i\frac{\Phi_{\rm L}}{2}}
\end{array}\right),
\label{self}
\end{eqnarray}
 and   
\begin{widetext}
\begin{eqnarray}
V_{\rm R}=\left(\begin{array}{cccc}
v_{\rm R}e^{i(\frac{\Phi_{\rm R}}{2}-\phi_{\rm RS})}  &  0 & 0 & 0\\
0  & -v_{\rm R}e^{-i(\frac{\Phi_{\rm R}}{2}+\phi_{\rm RS})} &  0 & 0\\
0  &  0 & v_{\rm R}e^{i(\frac{\Phi_{\rm R}}{2}+\phi_{\rm RS})}  &   0 \\
0  &  0 & 0 &  -v_{\rm R}e^{-i(\frac{\Phi_{\rm R}}{2}-\phi_{\rm RS})}
\end{array}\right).
\label{self}
\end{eqnarray}
\end{widetext}
Here, $t_{\rm L(R)}$ is the tunneling matrix and $V_{\rm L(R)}$ represents the coupling between the left (right) superconductor and the QD.
The Keldysh retarded Green's function can be expressed in a block form like
\begin{eqnarray}
g^{\rm r}=\left(\begin{array}{ccc}
g^{\rm r}_{\rm LL}  &  0 & 0 \\
0  & g^{\rm r}_{\rm RR} &  0 \\
0  &  0 & g^{\rm r}_{\rm dd}   
\end{array}\right),
\label{self}
\end{eqnarray}
where
\begin{eqnarray}
g^{\rm r}_{\alpha\alpha}=\left(\begin{array}{ccccc}
g^{\rm r}_{\alpha\alpha,\epsilon_{\uparrow},\epsilon_{\uparrow}}  &  g^{\rm r}_{\alpha\alpha,\epsilon_{\uparrow},-\epsilon_{\uparrow}} & 0 & 0  \\
g^{\rm r}_{\alpha\alpha,-\epsilon_{\uparrow},\epsilon_{\uparrow}} & g^{\rm r}_{\alpha\alpha,-\epsilon_{\uparrow},-\epsilon_{\uparrow}}  &  0 & 0 \\
0  &  0 & g^{\rm r}_{\alpha\alpha,\epsilon_{\downarrow},\epsilon_{\downarrow}} & g^{\rm r}_{\alpha\alpha,\epsilon_{\downarrow},-\epsilon_{\downarrow}}  \\
0  &  0 & g^{\rm r}_{\alpha\alpha,-\epsilon_{\downarrow},\epsilon_{\downarrow}} & g^{\rm r}_{\alpha\alpha,-\epsilon_{\downarrow},-\epsilon_{\downarrow}} 
\end{array}\right)\nonumber \\
\label{self}
\end{eqnarray}
with $\alpha \in \rm L/\rm R$. The retarded Green's functions for the each individual leads are expressed as, $g^{\rm r}_{\alpha\alpha,\epsilon\sigma,\epsilon\sigma}=-i\pi\rho\rho_{\alpha}$ and
$g^{\rm r}_{\alpha\alpha,\pm \epsilon\sigma,\mp\epsilon \sigma}=-i\pi\rho\rho_{\alpha}\sigma\Delta_{0}/(\epsilon+i\eta^{+})$. Here, the density of states of the QD is taken as $\rho=1$ and the density of states of the leads are given by
\begin{figure}
    \includegraphics[scale=0.75]{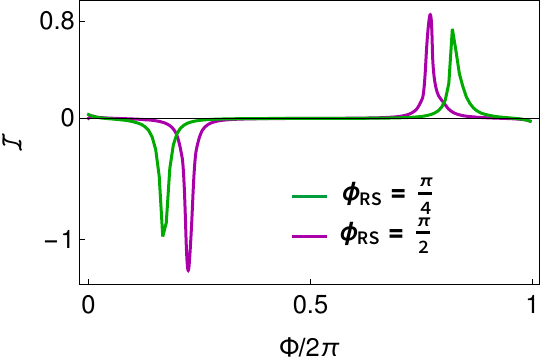}
    \caption{Josephson current $I$ in units of $e\Delta_{0}$ as a function of the phase difference for the asymmetric JJ. }
    \label{fig:RC_AL}
    \end{figure}
\beq
\rho_{\alpha}=\begin{cases}
\frac{|\epsilon|}{\sqrt{\epsilon^{2}-\Delta^{2}_{0}}}  & |\epsilon|>\Delta_{0}\\
\frac{-i\epsilon}{\sqrt{\Delta^{2}_{0}-\epsilon^{2}}} & |\epsilon|<\Delta_{0}.
\end{cases}
\eeq
\begin{figure}[th]
\includegraphics[scale=0.75]{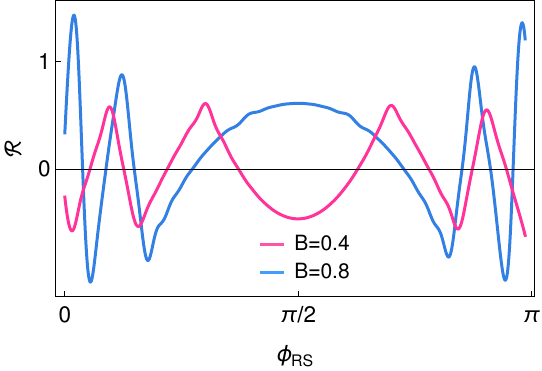} 
    \caption{RC ($\mathcal{R}\%$) vs. Rashba phase in the absence of chirality.}
    \label{fig:RClamdazero}
    \end{figure}

The Green's function of the uncoupled QD is calculated as:
\begin{widetext}
\begin{eqnarray}
g^{\rm r}_{\rm dd}=\left(\begin{array}{cccc}
\frac{1}{\epsilon+i\eta^{+}-H_{\rm QD,\uparrow\uparrow}}  & 0 & 0 & 0 \\
0  & \frac{1}{-\epsilon+i\eta^{+}-H_{\rm QD,\uparrow\uparrow}} & 0 & 0\\
0 & 0 & \frac{1}{\epsilon+i\eta^{+}-H_{\rm QD,\downarrow\downarrow}}  & 0 \\
0 & 0 & 0  & \frac{1}{-\epsilon+i\eta^{+}-H_{\rm QD\downarrow\downarrow}} 
\end{array}\right).
\label{self}
\end{eqnarray}
\end{widetext}

Using the Keldysh retarded Green's function $g^{\rm r}$ of the uncoupled system and the individual coupling matrices ($V_{\rm L}, V_{\rm R}, t_{\rm LR}$), the self-energy of the system $\Sigma^{\rm r}$ is calculated using the Dyson's equation of motion. Further, the Keldysh lesser Green's function $G^{<}$  is computed numerically using the fluctuation-dissipation theorem following $G^{<}(\epsilon)=-f(\epsilon)(G^{\rm r}-G^a)$, where $G^{\rm a}=[G^{r}]^{\dagger}$. Finally, we calculate the Josephson current self-consistently using the matrix components of the Green's function.

\section{Effect of asymmetric leads}\label{apnd2:asymmetric}
In this section, we study the effect of asymmetric superconducting leads on the current profiles.

To impose the asymmetry between the two leads in our JJ, we consider $\Delta_{\rm L}=1$ and $\Delta_{\rm R}=0.6$ and plot the current in terms of the phase difference in Fig.~\ref{fig:RC_AL}. We observe that the positive and the negative critical currents are different from each other. Comparing with Fig.~\ref{fig:Jvsphi_R}, it is clearly visible that for the asymmetric leads, the non-reciprocity in the Josephson current increases. Also, the negative current is now higher than the positive current unlike the situation for the symmetric leads. We have also checked (not shown to avoid increasing number of figures) that the corresponding RC of our diode also increases when the leads are asymmetric compared to that for symmetric case.

\section{Derivation of current expression}\label{apnd:currexpression}
In this section, we present an analytical expression for the Josephson current in the presence of RSOI. From Eq.\,\eqref{Eq:H_eff} we find the effect of Rashba interaction is included in the tunneling Hamiltonian. Therefore, using this effective Hamiltonian we calculate an approximate analytical expression for the Josephson current in our model.

The current can be defined in terms of the Free energy $F$ of the system as~\cite{Rasmussen2016, Davydova2022},
\beq
   I=\frac{2e}{\hbar}\frac{\partial F}{\partial \Phi}
   \eeq
where $e$ is the electronic charge and it is set as $e=1$. It can also expressed in terms of the partition function $Z$ as
\beq
    I =\frac{2e}{\hbar}\frac{\partial (-T \ln Z)}{\partial \Phi}.
    \eeq
Now explicitly writing the partition function of the junction, 
we derive the current as
\begin{eqnarray}  
I&=&-\frac{2e}{\hbar}\frac{1}{\beta}\frac{\partial (\ln \rm Tr[\exp^{-\beta \tilde{H}}])}{\partial \Phi} \non\\
& \simeq &v_{\rm L} \exp^{-i \frac{\phi_{\rm RS}}{2}} \sin\left({\frac{\Phi+\phi_{\rm RS}}{2}}\right)+v_{\rm L}\exp^{i \frac{\phi_{\rm RS}}{2}} \sin\left({\frac{\Phi-\phi_{\rm RS}}{2}}\right).  \non \\
\end{eqnarray}
\begin{figure}[ht!]
    \includegraphics[scale=0.67]{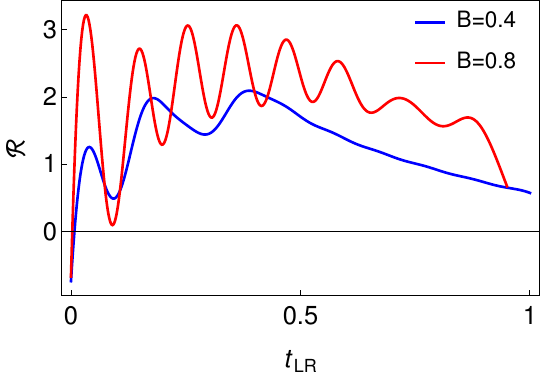}
    \caption{RC ($\mathcal{R}\%$) vs. hopping integral $t_{\rm LR}$  for $\phi_{RS}=\pi/4$.}
    \label{fig:Qvst}
\end{figure}
Collecting the real part of the above expression we obtain the approximate relation for the CPR mentioned in the main text.

\section{Rectification in absence of chirality}\label{apnd3:zerochirality}
In our QD-based JJ, we consider the presence of RSOI which eventually breaks the IS. The QD is modelled to have the chirality which also breaks the symmetry. To obtain the JDE, the chirality is not any necessary condition when RSOI is present. For the confirmation, we present the behavior of RC as a function of Rashba phase in the absence of $\lambda$ for different magnetic fields in Fig.~\ref{fig:RClamdazero}. The qualitative nature of the RC with the variation of RSOI is quite similar to that in the presence of chirality. Comparing with Fig.~\ref{fig:QvsR_B}, we find that the magnitude of RC is higher in the presence of the chirality. However, the sign changing phenomena is more prominent in the absence of the chirality. The sign of RC changes from the positive to the negative or vice versa with the change in Rashba phase. Thus, the sign and magnitude of the RC in our JD is sensitive to the chirality.
\section{Effect of direct coupling} \label{apnd4:directcoupling}
A direct coupling between the leads helps to tackle the effect of the RSOI analytically. In experimental fabrication, the coupling between the leads could be changed by forming the junction potential. Therefore, it is a controllable parameter in practice~\cite{Miller2003, Mayer2020}. In the main text, we discuss all results for a particular value of the direct coupling strength. Now, we show its impact on the RC of our JD in Fig.~\ref{fig:Qvst}. We find that the RC is oscillatory with respect to $t_{\rm LR}$ at a constant RSOI in the presence of the finite magnetic field. The direct coupling essentially plays the role of an extra-channel in the QD junction. Tuning the strength effectively leads to the controlling the extra channel which further modifies the tunneling phenomena and results in the oscillations. The oscillations increases with the increase in the magnetic field. 
\end{appendix}

\bibliography{bibfile}{}
\end{document}